%% file: rdt.tex
\input phyzzx
\input maggiemac
\hoffset=0.375in
\overfullrule=0pt

\def\max{{\rm max}}
\def\min{{\rm min}}
\def\lim{{\rm lim}}

\def\pc{{\rm pc}}
\def\kms{{\rm km}\,{\rm s}^{-1}}

\twelvepoint
\font\bigfont=cmr17
\centerline{\bigfont Femtolens Imaging of a Quasar Central Engine}
\smallskip
\centerline{\bigfont Using a Dwarf Star Telescope}
\bigskip
\centerline{{\bf Andrew Gould}\footnote{1}{Alfred P.\ Sloan Foundation Fellow}}
\smallskip
\centerline{and}
\smallskip
\centerline{\bf B.\ Scott Gaudi}
\smallskip
\centerline{Dept of Astronomy, Ohio State University, Columbus, OH 43210}
\smallskip
\centerline{e-mail gould@payne.mps.ohio-state.edu, 
gaudi@payne.mps.ohio-state.edu}
\bigskip
\centerline{\bf Abstract}
\singlespace 

We show that it is possible to image the structure of a distant quasar on
scales of $\sim 1\,$AU by constructing a telescope which uses a nearby dwarf 
star as its ``primary lens'' together with a satellite-borne ``secondary''.  
The image produced by the primary is magnified by $\sim 10^5$ in one direction 
but is contracted by 0.5 in the other, and therefore contains highly degenerate
one-dimensional information about the two-dimensional source.  We discuss 
various methods for extracting information about the second dimension including
``femtolens interferometry'' where one measures the interference between 
different parts of the one-dimensional image with each other.  Assuming that 
the satellite could be dispatched to a position along a star-quasar line of 
sight at a distance $r$ from the Sun, the nearest available dwarf-star primary 
is likely to be at $\sim 15\,\pc\,(r/40\,\rm AU)^{-2}$.  The secondary should 
consist of a one-dimensional array of mirrors extending $\sim 700\,$m to 
achieve 1 AU resolution, or $\sim 100\,$m to achieve 4 AU resolution.

\bigskip
Subject Headings: gravitational lensing -- instrumentation: interferometers 
-- quasars: general
\smallskip
\centerline{submitted to {\it The Astrophysical Journal Letters}: 
June 3, 1996}
\centerline{Preprint: OSU-TA-13/96}

\endpage
\normalspace
\chapter{Introduction}

	Gravitational microlensing by Galactic stars (and possibly other
objects) has been detected in more than 100 events seen toward the Galactic
bulge by MACHO (Alcock et al.\ 1996), OGLE (Udalski et al.\ 1994), and
DUO (Alard 1996).  Generally, these stellar lenses are believed to be
several kpc from the Sun, either in the Galactic bulge or in distant regions
of the Galactic disk.

	There have been several suggestions to make use of more nearby stars
as lenses.  Nearby stars have high proper motions and relatively large
Einstein rings.  One could predict when they would lens more distant stars,
measure the Einstein ring and so determine the mass of the lensing star
(Paczy\'nski 1995).  Even if the projected separation between the nearby
and distant stars is many Einstein radii, the former will
still deflect the light of the latter. This effect 
could be measured using interferometry and would again yield a mass
measurement (Miralda-Escud\'e 1996).  Extremely nearby objects could
conceivably be detected by diffractive lensing flashes (Labeyrie 1994).

	Here we propose a more ambitious use of a nearby star as the primary
lens of a telescope whose purpose would be to resolve the central engine of a 
quasar.  Since quasars lie at an angular-diameter distance 
$D_Q\sim 10^9\,$pc, and since their 
central engines are believed to have dimensions ${\cal O}$(AU), such a 
telescope would require an effective resolution of $10^{-9}$ arcsec,
roughly the equivalent of an optical interferometer with an Earth-Moon
baseline.  In general, previous 
suggestions for the use of nearby stars have assumed that one must wait for 
the lens and source to line up with the Earth.  The probability for the
required alignment with a quasar is negligibly small.  However, if Mohammed
will not go to the mountain ...

\chapter{Characteristics of the Primary Lens}

	To construct the telescope, there must be a dwarf star aligned 
with a quasar as seen from {\it somewhere} in the solar system.  For 
definiteness, we define ``somewhere'' as being within a distance $r=40\,$AU 
of the Sun.  For a star at distance $D$, the quasar must then lie within
an angle $r/D$ of the star (as seen from the Sun).  The probability of
such an occurrence is $\pi(r/D)^2 N_Q$, where $N_Q= 200\,\rm deg^{-2}$ is the 
angular density of quasars brighter than $B=22$ (Hartwick \& Schade 1990).
The expected number of such alignments by dwarfs within a distance 
$D_\lim$ of the Sun is then 
$$\int_0^{D_\lim} d D\, 4\pi D^2 \pi \biggl({r\over D}\biggr)^2 n N_Q 
= 4\pi^2 r^2 \,D_\lim n N_Q \sim {D_\lim\over 15\,\rm pc},\eqn\numest$$ 
where we have assumed a local density of dwarfs of $n=0.07\,\rm pc^{-3}$
(Gould, Bahcall, \& Flynn 1996 and references therein).
Even if no such dwarfs are present at a given time, a statistically
independent sample will be available after a duration $r/v\sim 5\,$years,
where we have adopted $v\sim 40\,\kms$ as the typical transverse speed of
disk stars.  

	We now assume that such a star has been identified at a distance
$D\sim 15\,$pc and mass $M\sim 0.4 M_\odot$.  We further assume that a 
satellite has been equipped with the remaining
optics that are required for construction of the telescope and that it
has been dispatched to a position along the dwarf-quasar line of sight.  
Note that the satellite must not only be in the correct location, but must 
also be moving with the same transverse velocity as the dwarf star 
($\sim 40\,\kms$) so that the alignment with the quasar is maintained.  

	The first job of the satellite will be to measure the optical 
properties of the primary lens.  The distance to the lens will be very 
accurately known from parallax measurements.  Hence the mass can be 
determined from the angular size of the Einstein ring 
$\theta_e=(4 G M/c^2 D)^{1/2}$.  Since
the quasar is at a cosmological distance, its light will be affected
by a gravitational shear due to mass distributed near the line of sight.
This shear combines with the point-mass lens (the dwarf star) to produce
a Chang-Refsdal lens (Chang \& Refsdal 1979).  Such lenses are completely
described by the angular size of the Einstein ring, $\theta_e$, the magnitude
of the shear, $\gamma$, and the orientation of the shear.  The magnitude
of the shear will not be known {\it a priori}, but based on initial 
measurements or possibly upper limits (Villumsen 1996), we assume $\gamma\sim
1\%$.  For $\gamma\ll 1$ it is easy to show (using e.g.\ the formalism of
Gould \& Loeb 1992) that the lens has a diamond-shaped caustic with an angular 
distance between opposite cusps $\sim 4\gamma\theta_e$.  Hence one could
measure $\gamma$ by moving the spacecraft so as first to align the quasar
with one of the cusps and then measuring the distance to the opposite
cusp.  For our adopted parameters, $\theta_e\sim 10\,$mas and the distance
between cusps is $4\gamma D\theta_e= 6\times 10^{-3}$ AU.  The orientation of 
the caustic is along the line connecting the star and the quasar
image when the quasar is at the cusp.

\chapter{Characteristics of the Satellite}

	To resolve the central engine of the quasar, the satellite must
situate itself so that the center of the quasar lies inside the caustic
and very near the cusp.  Let $(\xi,\eta)$ be the coordinates of source
plane in units of the Einstein ring.  One may equally well think of these
as coordinates of the observer plane with corresponding physical distances
$(D\theta_e\xi,D\theta_e\eta)$.  Let $\Delta \xi$ be the distance along
the $\xi$ axis from the cusp at $\xi\sim 2\gamma$.  One finds that the
caustic in the neighborhood of the cusp is reasonably well described by
$\eta^2=(\Delta\xi)^3/4\gamma$.  Thus, a source with angular radius
$\rho\theta_e$ can come as close as $\Delta\xi\sim(4\gamma\rho^2)^{1/3}$
while still remaining inside the caustic.  For sources lying on the $\xi$ axis,
the magnification tensor of the brightest image is diagonal, with diagonal
components $(\Delta\xi^{-1},0.5)$.  Hence, a source of size 
$\rho\theta_e\sim AU/D_Q\sim 10^{-9}$ arcsec is stretched by a factor
$\sim 1.4\times 10^{5}$ and would therefore subtend an angle 
$\theta_r\sim140 \mu$as.

	If the quasar were observed in optical light $(\lambda\sim 0.5\,\mu$m),
an effective aperture $\lambda/\theta_r\sim 700\,$m would be required to 
resolve this
angle.  The required aperture is therefore smaller than the full width caustic,
$2D\rho\theta_e\sim 5\,$km, implying that the entire effective aperture would
fit well inside the caustic projected onto the observer plane.  This
is an essential ingredient for making the observations.  It is impractical
and unnecessary to fill the entire aperture with mirrors.  In fact, only one
dimension must be well sampled
because the quasar image is essentially one dimensional:
it is highly stretched in one direction and slightly contracted in the
other.  Even the image of the region $\sim 2000$ AU around the quasar would 
have a lateral extent of only $\sim 1\,\mu$as, far too small to be resolved.
Of course, it will be necessary to remove the light from the lensing star 
which will be separated from the quasar image by $\theta_e\sim 10^4\,\mu$as
and will typically subtend $\sim 300\,\mu$as. Most of this light could be
directed to a position well off the image center simply by tilting the
axis of the array $\sim 20^\circ$ relative to the critical curve.  There
would still be some contamination from the high-order fringes but much of
this could be removed provided that the width of the mirror array were a few 
tens of meters.

	To remain in the caustic over time, the satellite must
counter the acceleration due to the Sun at a distance of 40 AU, which is
$\sim 4\times 10^{-4}\,\rm cm\,s^{-2}$.
After 10 hours of such acceleration, the satellite
will have drifted $\sim 3\,$km and will require a boost of $\sim 20\,\rm cm
\,s^{-1}$ to get back on track.  The energy required for these repeated boosts
is negligible compared to that needed to achieve the initial velocity.
However, stabilization of the mirror system following the boosts might prove
to be a significant engineering challenge.

\chapter{Information in One and Two Dimensions}

\FIG\one{
Primary lens geometry.  Source lies entirely inside the caustic
and near one cusp.  It is therefore lensed into four images, three of which
are highly magnified and lie very close to the critical curve to the right.
Note that they are essentially one-dimensional.  The fourth image (to the 
left) is not highly magnified.  The inset shows a close up of the source
within the caustic.  For this illustrative example, the source is
stretched only by a factor $\sim 10^2$ compared to factors $\sim 10^5$
generally considered in this paper.
}
\topinsert
\mongofigure{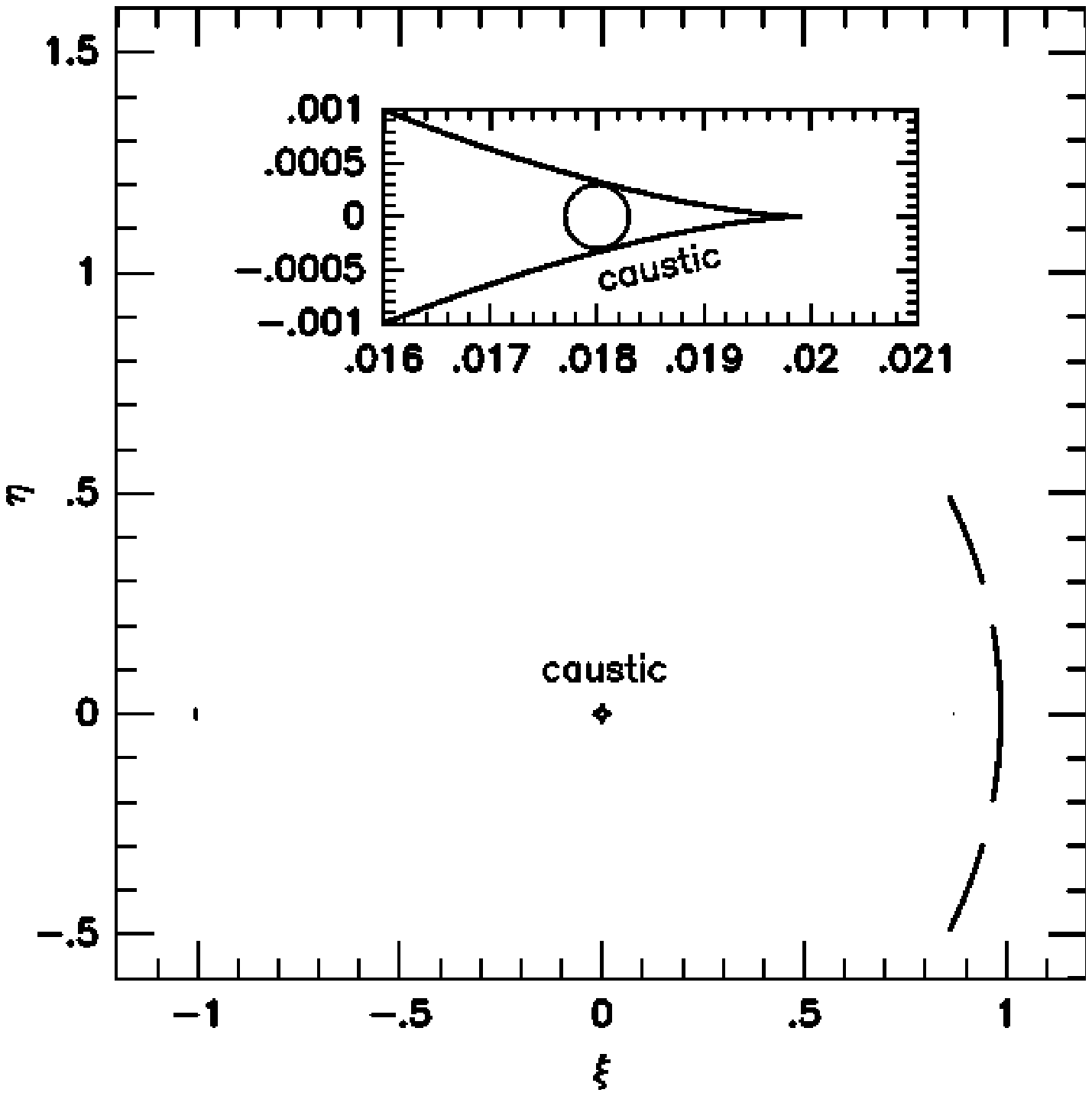}{6.4}{5.5}{6.4}{
Primary lens geometry.  Source lies entirely inside the caustic
and near one cusp.  It is therefore lensed into four images, three of which
are highly magnified and lie very close to the critical curve to the right.
Note that they are essentially one-dimensional.  The fourth image (to the 
left) is not highly magnified.  The inset shows a close up of the source
within the caustic.  For this illustrative example, the source is
stretched only by a factor $\sim 10^2$ compared to factors $\sim 10^5$
generally considered in this paper.
}
\endinsert

	The source regions inside the caustic are mapped into four images.  
For a point
source near the cusp, three of these images are highly 
magnified and lie close to the critical curve on the same side of lensing
star as the quasar.  The fourth image lies on the opposite side of the
star.  It is not highly magnified and will not be considered further.
For $\gamma\ll 1$, the critical curve deviates only slightly from the
Einstein ring of an isolated lens.  The central image is the brightest
and lies just outside the critical curve.  The other two images lie just
inside the critical curve and the sum of their magnifications is 
nearly equal to that of the central image.  Each image with magnification
$A$ is stretched by a factor $2A$ along the critical curve and compressed
by a factor 0.5 perpendicular to it.  See Figure \one.

\FIG\two{
Bands of image degeneracy.  The bold curve is the caustic within
which there are three images.  All three images lie in an almost 
one-dimensional structure close to the critical curve.  All the points in the
source between two adjacent solid lines are mapped into the same resolution 
element ($\sim 140\,\mu$as) along the critical-curve image.  The bands are
labeled 1, 2, 3, and 4.  Band 1 is marked with a dotted line.
The 10 AU circle at the lower right indicates the scale and distortion of the
figure which, for reasons of clarity, is stretched by a factor 10 in the 
vertical direction.  A smaller circle of radius 1 AU is marked ``central
engine''.  Note that it is resolved into two resolution elements.
}
\topinsert
\mongofigure{rdtfig2.ps}{6.4}{4.0}{6.4}{
Bands of image degeneracy.  The bold curve is the caustic within
which there are three images.  All three images lie in an almost 
one-dimensional structure close to the critical curve.  All the points in the
source between two adjacent solid lines are mapped into the same resolution 
element ($\sim 140\,\mu$as) along the critical-curve image.  The bands are
labeled 1, 2, 3, and 4.  Band 1 is marked with a dotted line.
The 10 AU circle at the lower right indicates the scale and distortion of the
figure which, for reasons of clarity, is stretched by a factor 10 in the 
vertical direction.  A smaller circle of radius 1 AU is marked ``central
engine''.  Note that it is resolved into two resolution elements.
}
\endinsert

	In principle, since each point of the source is mapped into three
unique points of the image, one can reconstruct the source from the image.
In fact, all three images are compressed into a one dimensional curve whose
angular width is approximately equal to that of the unlensed source. As
mentioned above, this means that a region of 2000 AU is compressed into
a curve of width $\sim 1\,\mu$as, far smaller than the resolution element.
Thus, in practice each resolution element (``point'') along the critical curve 
represents the sum of the contributions from an entire curve within the source.
These ``degeneracy curves'' are actually almost perfectly straight lines and 
are tangent to the caustic at exactly one point.  Since the resolution element
has finite width, the  ``degeneracy curves'' are in fact ``degeneracy bands''
whose width shrinks as it approaches the tangent point and then expands again.
See Figure \two.  

	The nature of this degeneracy can be understood by considering
a quasar with a central engine of radius $\sim 1\,$AU, but which is
emitting significant amounts of light over a radius $\sim 200\,$AU.  As 
discussed above, the central AU is imaged into a rectangular arc about
$100\,\mu$as$\times 10^{-3}\,\mu$as.  Other regions of the quasar of
intrinsic width 1 AU will
be mapped into the same $100\,\mu$as, also with width $10^{-3}\,\mu$as.
The relative contributions of these different regions to the observed light
will scale according to their surface brightnesses.  Thus, if the entire quasar
had the same surface brightness, the central engine would contribute only
$\sim 1/200$ of the light to this resolution element of the image.  
If surface brightness
fell inversely with radius, it would contribute a fraction $1/\ln(200)\sim
20\%$.

	There are several methods for breaking this degeneracy.  The first is
to make use of the three images.  A most spectacular example of this approach
is the recent resolution of an apparent ring galaxy using multiple images
produced by a cluster lens (Colley, Tyson, \& Turner 1996).  If there is
a significant excess brightness of the middle image due to a hot central 
engine, this excess will also be apparent at the locations of the two other 
images.

	There is also a second more powerful method.  As the satellite drifts
over the central regions of the quasar, the point in the source that is
nearest the cusp and which is therefore maximally magnified varies. 
The positions within the caustic structure of all other points in the source
also vary, but the change in their magnifications with position
is much smaller. 
Thus, one could map out the brightness of the inner regions.  The second
and third images would serve as a check on these measurements.  Of course,
this approach requires that the brightness profile of the quasar remain
relatively constant during the series of observations.

\chapter{Femtolens Interferometry}

	Still another method  is to look for interference effects between
images.  A number of authors have discussed the possibility of observing 
``femtolensing'', interference between the {\it integrated light} of two or 
more images
(Mandzhos 1981; Schneider \& Schmidt-Burgk 1985; Deguchi \& Watson 1986;
Peterson \& Falk 1991; Gould 1992; Ulmer \& Goodman 1995).  Here we
analyze the possible role of such interference in the {\it two-dimensional
reconstruction}
of the source.  The principle is relatively simple.  As shown in Figure
\two, each resolution element of the one-dimensional image along the critical 
curve corresponds to a band of variable width in the source plane.  
Consider the two bands labeled ``1'' and ``3''.
Suppose that 
the light falling on the two 
corresponding resolution elements is brought together and then dispersed in a 
spectrograph.  Most of the light in each resolution element comes from regions 
of the source that are unrelated to the regions that generate the light 
entering the other element.  This light does not suffer any effects
of interference.  However, the light coming from the small shaded region 
where the two sets of curves cross arrives at the two images ``1'' and ``3'' 
at times
$t_1$ and $t_3$ and therefore suffers interference according to a time lag
$\Delta t_{1,3}=t_1-t_3$.  Hence, there will be an oscillation in the 
spectrum with frequency spacing $\Delta f_{1,3}= (\Delta t_{1,3})^{-1}$.  
Such an oscillation is observable provided the spectrograph has a resolving 
power $R\gsim c t_{1,3}/\lambda$ and, of course, provided it has sufficient 
amplitude to be seen above the noise of the remaining photons that do not 
participate in the fluctuation.  

\FIG\three{
Overlap between source resolution elements ``1'' and ``3'' from Fig.\ \two\ 
({\it bold dashed quadrangle}) subdivided into ten zones of equal
time delay between the two images ({\it solid up-right diagonal curves}).  
The total light in each zone can be determined by Fourier analysis of the
interference resolution elments ``1'' and ``3''.  The overlap
between source resolution elements ``1'' and ``--4'' ({\it bold quadrangle})
is also divided into
ten zones of equal delay between images ``1'' and ``--4'' ({\it solid up-left
diagonal curves}).  A similar (approximately horizontal) set of contours could
also be drawn for the delay structure between images ``3'' and ``--4'' 
but is not shown to avoid clutter.  These measurements could produce resolution
that is substantially finer than the naive result indicated in Fig.\ \two.
}
\topinsert
\mongofigure{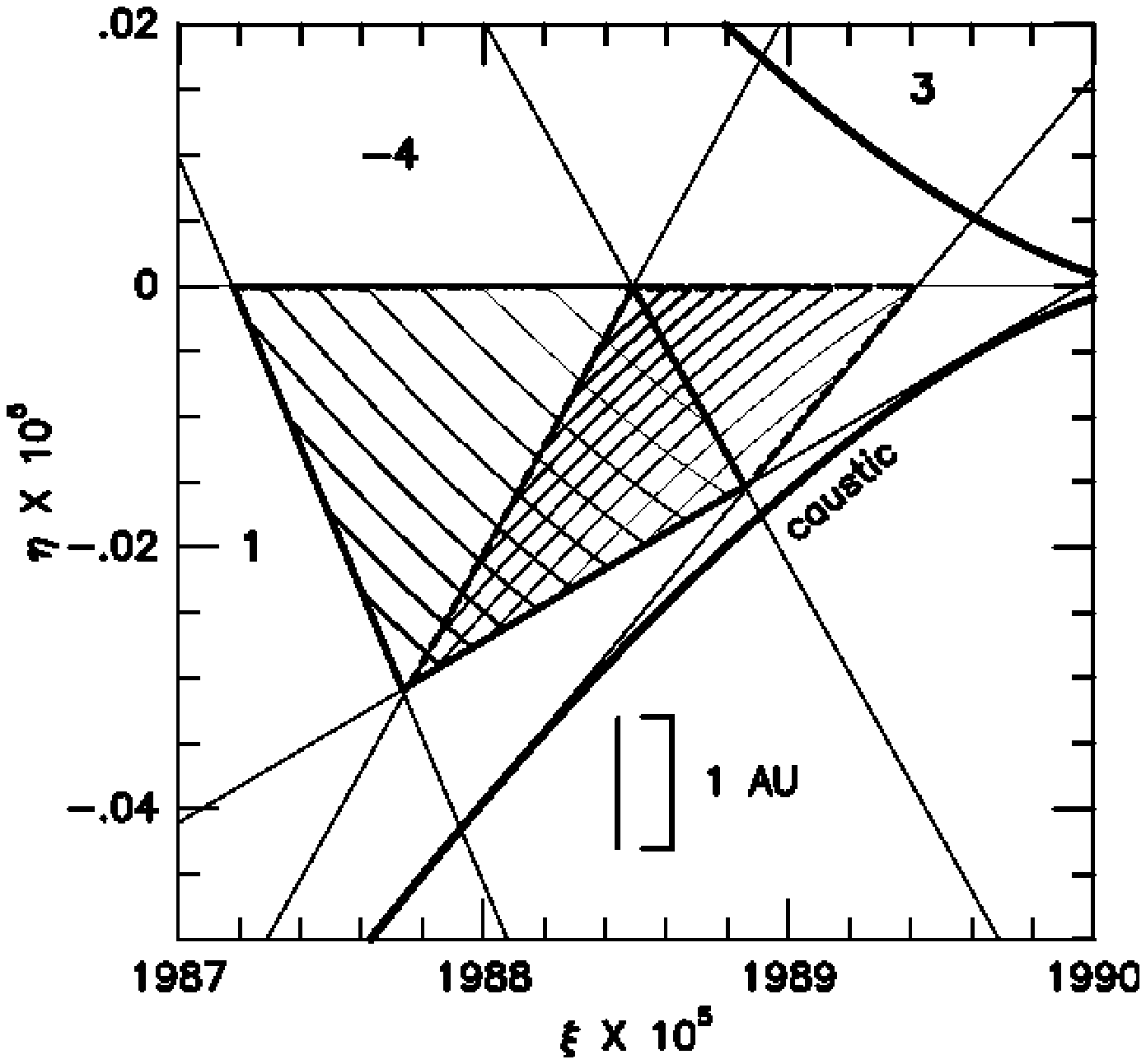}{6.4}{5.5}{6.4}{
Overlap between source resolution elements ``1'' and ``3'' from Fig.\ \two\ 
({\it bold dashed quadrangle}) subdivided into ten zones of equal
time delay between the two images ({\it solid up-right diagonal curves}).  
The total light in each zone can be determined by Fourier analysis of the
interference resolution elments ``1'' and ``3''.  The overlap
between source resolution elements ``1'' and ``--4'' ({\it bold quadrangle})
is also divided into
ten zones of equal delay between images ``1'' and ``--4'' ({\it solid up-left
diagonal curves}).  A similar (approximately horizontal) set of contours could
also be drawn for the delay structure between images ``3'' and ``--4'' 
but is not shown to avoid clutter.  These measurements could produce resolution
that is substantially finer than the naive result indicated in Fig.\ \two.
}
\endinsert

	It is actually possible to substantially further improve the 
resolution.  In the previous treatment, we imagined that the time delay 
between the ``1'' and ``3'' images is constant over the entire overlap region 
shown in Figure \two.  If this were so, the interference term would scale with
frequency $f$ as $\sin^2(2\pi f/f_{1,3})$.  In fact, the time delay is a
function of position within the patch, ranging from $\Delta t_{1,3,\min}=
2\lambda/c$ to $\Delta t_{1,3,\max}=33\lambda/c$, where $\lambda=0.5\mu$m.  
In Figure \three, we
show a blow-up of the overlap region with 10 contours of constant time delay
between the images in resolution elements ``1'' and ``3''.  By taking the
Fourier transform of the interference pattern produced by combining these
two resolution elements, one could determine how much of the source light
lay between pairs of consecutive contours.  Of course, the choice of 10 
contours was arbitrary. One could in principle 
resolve the region into up to the
diffraction limit of $\sim 30$ bands at which point the adjacent time delays
would differ by only $\sim \lambda/c$.  There is also a third image.  For
the left hand side of the overlap region this image lies in resolution element
``--4'', and for the right hand side it lies in element ``--3''.  We focus on 
the first.  One can now repeat the same procedure for the 
interference patterns generated by the time delays $\Delta t_{1,-4}$
(see Fig.\ \three) and $\Delta t_{3,-4}$, thereby further constraining the 
structure within the patch.  Note that while the one-dimensional resolution is
limited only by the diffraction limit, the true two-dimensional resolution is 
not as small
as the grid shown in Figure \three, since there are 100 such patches
in an overlap region (assuming 10 contours in each direction), but only
30 pieces of information from the three sets of contours.  Nevertheless, it
would be possible to track isolated blobs having the size of one of the
patches shown in Figure \three, and thus in this restricted sense the figure
is indicative of the effective resolution.

\chapter{Photons}

	It would be premature to attempt a detailed estimate of the 
signal-to-noise ratios in this proposed experiment.  However, it is legitimate
to ask whether these ratios are closer to $10^2$ or $10^{-2}$.  Suppose that
the quasar magnitude is $V=22$ and that 10\% of the quasar light comes out
in the {\it annulus} containing the shaded overlap region in Figure \two.
The fraction of the total light produced by the overlap region is then
$\sim 2\times 10^{-4}$.  The typical magnifications in this region
are $\sim 3\times 10^4$, i.e, each image has a (magnified) apparent magnitude
$V\sim 20$.  If we suppose that the total area of the telescope
were $20\,\rm m^2$, then the number of photons collected from this region
during a 1 hour exposure would be ${\cal O}(10^6)$.  If there were $\sim 10$
times more photons from non-overlap regions, the signal-to-noise ratio would
be $\sim 300$, an ample number to measure not only the total
flux from the region,
but substantial substructure as well.  Thus it would be possible to follow
the detailed evolution of the quasar on time scales comparable to the light
crossing time.

{\bf Acknowledgements}:  We thank D.\ DePoy and G.\ Newsom for several
useful suggestions.
This work was supported in part by grant AST 94-20746 from the NSF.
\bigskip
\Ref\alard{Alard, C.\ 1996, in Proc. IAU Symp.\ 173 (Eds.\ C.\ S.\ Kochanek, 
J.\ N.\ Hewitt), in press (Kluwer Academic Publishers)}
\Ref\Alcock{Alcock, C., et al.\ 1996, ApJ, submitted}
\Ref\cr{Chang, K., \& Refsdal, S.\ 1979, Nature, 282, 561}
\Ref\ctt{Colley, W.\ N., Tyson, J.\ A., \& Turner, E.\ L.\ 1996, ApJ, 461, L83}
\Ref\dw{Deguchi, S., \& Watson, W.\ D.\ 1986, ApJ, 307, 30}
\Ref\gou{Gould, A.\ 1992, ApJ, 386, L5}
\Ref\gbf{Gould, A., Bahcall, J.\ N., \& Flynn 1996, ApJ, 465, 000}
\Ref\gl{Gould, A., \& Loeb, A.\ 1992, ApJ, 396, 104}
\Ref\hart{Hartwick, R.\ D.\ A.\ \& Schade, D.\ 1990, AARA, 28, 437}
\Ref\Lab{Labeyrie, A.\ 1994, A\&A, 284, 689}
\Ref\Man{Mandzhos, A.\ V.\ 1981, Soviet Astron.\ Lett., 7, 213}
\Ref\Mir{Miralda-Escud\'e, J., 1996, ApJL, submitted}
\Ref\Pac{Paczy\'nski, B.\ 1995, Acta Astr., 45, 345}
\Ref\PF{Peterson, J.\ B., \& Falk, T.\ 1991, ApJ, 374, L5}
\Ref\ssb{Schneider, P., \& Schmidt-Burgk, J.\ 1985, A\&A, 148, 369}
\Ref\oglea{Udalski, A., et al.\ 
1994, Acta Astronomica 44, 165}
\Ref\UM{Ulmer, A., \& Goodman, J.\ 1995, ApJ, 442, 67}
\Ref\Villumsen{Villumsen, J.\ V.\ 1996, MNRAS, submitted}
\refout
\endpage
\bye

%% file: maggiemac.tex
\newcount\mongocount
\mongocount=1
\def\Figure#1#2#3{
      \vbox to #3in{\hsize=#2in
        \vfil
         \includegraphics{#1}
    }
}
\def\figcap#1#2{
\vtop{\tenpoint\singlespace
\hsize=#1in\smallskip\noindent Figure\ \ \the\mongocount.\ \  #2
\global\advance\mongocount by 1\bigskip}}
\def\mongofigure#1#2#3#4#5{\centerline{\Figure{#1}{#2}{#3}
\figcap{#4}{#5}}}